\documentclass[prd,twocolumn,showpacs,amsmath,amssymb,nofootinbib]{revtex4-2}

\usepackage{epsf}
\usepackage{graphicx}
\usepackage{epsfig}
\usepackage{xcolor}
\usepackage{subfigure}
\usepackage{pstricks}
\usepackage{pst-node}
\usepackage{rotating}
\usepackage{graphics}
\usepackage{latexsym}
\usepackage{dcolumn}
\usepackage{bm}
\usepackage{times}
\usepackage{indentfirst}
\usepackage{float}
\usepackage{overpic}
\usepackage{amsmath}
\usepackage{appendix}
\usepackage{upgreek}
\usepackage{chngcntr}
\usepackage{amssymb}%
\usepackage{pifont}%
\usepackage[top=0.5in,bottom=0.8in,left=0.9in,right=0.9in]{geometry}
\usepackage{subfigure}
\usepackage{lineno}

\usepackage{multirow}
\usepackage{makecell}

\usepackage{hyperref}
\hypersetup{
	colorlinks=true,
	linkcolor=blue,
	filecolor=blue,
	urlcolor=blue,
	citecolor=blue,
}

\newcommand{\ppbpi}{p\bar{p}{\pi}^{0}}

\newcommand{\EE}{e^+e^-}

\newcommand{\jpsi}{J/\psi}
\newcommand{\ar}{\rightarrow}
		
		\newcommand{\bfg}{\begin{figure}}
			\newcommand{\efg}{\end{figure}}
		\newcommand{\bitm}{\begin{itemize}}
			\newcommand{\eitm}{\end{itemize}}
		\newcommand{\bnum}{\begin{enumerate}}
			\newcommand{\enum}{\end{enumerate}}
		\newcommand{\btbl}{\begin{table}}
			\newcommand{\etbl}{\end{table}}
		\newcommand{\btbu}{\begin{tabular}}
			\newcommand{\etbu}{\end{tabular}}
		\newcommand{\bcl}{\begin{center}}
			\newcommand{\ecl}{\end{center}}
		
		\newcommand{\beq}{\begin{equation}}
			\newcommand{\eeq}{\end{equation}}
		\newcommand{\beqr}{\begin{eqnarray}}
			\newcommand{\eeqr}{\end{eqnarray}}
\begin{document}
	
\title{\boldmath Measurement of the $\EE\ar\ppbpi$ cross section at $\sqrt{s}=2.1000-3.0800$ GeV}
	
\author{
M.~Ablikim$^{1}$, M.~N.~Achasov$^{4,c}$, P.~Adlarson$^{75}$, O.~Afedulidis$^{3}$, X.~C.~Ai$^{80}$, R.~Aliberti$^{35}$, A.~Amoroso$^{74A,74C}$, Q.~An$^{71,58,a}$, Y.~Bai$^{57}$, O.~Bakina$^{36}$
% ...
\\
\vspace{0.2cm}
(BESIII Collaboration)\\
\vspace{0.2cm} {\it
$^{1}$ Institute of High Energy Physics, Beijing 100049, People's Republic of China\\
% ...
$^{3}$ Bochum  Ruhr-University, D-44780 Bochum, Germany\\
$^{4}$ Budker Institute of Nuclear Physics SB RAS (BINP), Novosibirsk 630090, Russia\\
% ...
$^{35}$ Johannes Gutenberg University of Mainz, Johann-Joachim-Becher-Weg 45, D-55099 Mainz, Germany\\
$^{36}$ Joint Institute for Nuclear Research, 141980 Dubna, Moscow region, Russia\\
% ...
$^{57}$ Southeast University, Nanjing 211100, People's Republic of China\\
$^{58}$ State Key Laboratory of Particle Detection and Electronics, Beijing 100049, Hefei 230026, People's Republic of China\\
% ...
$^{71}$ University of Science and Technology of China, Hefei 230026, People's Republic of China\\
% ...
$^{74}$ University of Turin and INFN, (A)University of Turin, I-10125, Turin, Italy; (B)University of Eastern Piedmont, I-15121, Alessandria, Italy; (C)INFN, I-10125, Turin, Italy\\
$^{75}$ Uppsala University, Box 516, SE-75120 Uppsala, Sweden\\
% ...
$^{80}$ Zhengzhou University, Zhengzhou 450001, People's Republic of China\\
\vspace{0.2cm}
$^{a}$ Deceased\\
% ...
$^{c}$ Also at the Novosibirsk State University, Novosibirsk, 630090, Russia\\
% ...
}
}

\date{\today}

\renewcommand{\abstractname}{}
\begin{abstract}
The process $e^{+}e^{-}\to p\bar{p}\pi^{0}$ is studied at 20 center-of-mass energies ranging from 2.1000 to 3.0800 GeV using 636.8~pb$^{-1}$ of data collected with the BESIII detector operating at the BEPCII collider.
The Born cross sections for $e^{+}e^{-}\to p\bar{p}\pi^{0}$ are measured with high precision.  
Since the lowest center-of-mass energy, 2.1000~GeV, is less than 90~MeV above the $p\bar{p}\pi^0$ energy threshold, we can probe the threshold behavior for this reaction.
However, no anomalous threshold enhancement is found in the cross sections for $e^{+}e^{-}\to p\bar{p}\pi^{0}$.
\end{abstract}	
	
\maketitle

\section{INTRODUCTION} \label{}
Studies of threshold enhancements in the pair production of ground-state baryons in the spin-$\frac{1}{2}$ $SU\text{(3)}$ octet provide important information about the interactions and fundamental structure of particles.
Significant threshold enhancements have previously been observed in the cross sections for the processes $e^{+}e^{-}\to p\bar{p}$~\cite{ppb2015,ppb2020,ppb2019,ppb2021,ppb2021r}, $e^{+}e^{-}\to{\Lambda}_{c}\bar{{\Lambda}}_{c}$~\cite{lamclamc2018, lamclamc2023}, $e^{+}e^{-}\to\Lambda\bar{\Lambda}$~\cite{lamlam2018,lamlam2023}, etc. 
A common feature in the measured cross sections for many of these processes is a plateau starting from the baryon pair production threshold~\cite{overall}.
For asymmetric baryon pair, a non-zero Born cross section for the process $e^{+}e^{-}\to \Lambda\bar{\Sigma}^{0}+c.c.$ at center-of-mass~(c.m.) energy~($\sqrt{s}$) of 2.3094~GeV, which is about 1~MeV above the production threshold, is observed with a statistical significance of more than five standard deviations~\cite{LambdaSigma}.
Within the process $e^{+}e^{-}\to p\bar{p}\pi^{0}$, the subprocess $e^{+}e^{-}\to \bar{p}{N}^{*}+\text{c.c.}$ or $e^{+}e^{-}\to \bar{p}{\Delta}^{*}+\text{c.c.}$, where $N^{*}$ and ${\Delta}^{*}$ couple to $p\pi^{0}$, could potentially show a similar behavior. 

The Born cross sections of $e^{+}e^{-}\to p\bar{p}\pi^{0}$ have previously been investigated in the charmonium region at BESIII.
For example, around the $\psi(3770)$ resonance the $e^+e^-$ cross sections have been measured and the Born cross section of $\psi(3770)\to p\bar{p}{\pi}^{0}$ has been extracted considering interference between resonant and continuum production amplitudes~\cite{3770ppbpi}.
Also, at $\sqrt{s}$ ranging from 4.008 to 4.600~GeV this process has been studied to explore the properties of charmonium states~\cite{4260ppbpi}.
However, measurements of the Born cross sections at lower $\sqrt{s}$, especially near the $p\bar{p}{\pi}^{0}$ production energy threshold, remain insufficient.
Accurately characterizing the cross sections for $e^{+}e^{-}\to p\bar{p}\pi^{0}$ in the lower energy domain poses significant challenges and tests for perturbative QCD, especially near the threshold.
In addition, these Born cross sections can be used to estimate the cross sections for $p\bar{p}\to {e}^{+}{e}^{-}{\pi}^{0}$.  These in turn can be used to measure the proton form factor in the unphysical region~\cite{pppiee}, which is a goal for the future anti-Proton ANnihilations DArmstadt~($\rm \overline{P}$ANDA) experiment at the Facility of Anti-proton and Ion Research~(FAIR) in Germany~\cite{pandaphy}. 

In this paper, we report measurements of the Born cross sections for the process $e^{+}e^{-}\to p\bar{p}\pi^{0}$, obtained by analyzing ${e}^{+}{e}^{-}$ collision data collected with the BESIII detector at $\sqrt{s}$ ranging from 2.1000 to 3.0800~GeV with a total integrated luminosity of 636.8~pb$^{-1}$.

\section{THE BESIII DETECTOR AND DATA SAMPLES}
The BESIII detector~\cite{BESIII} records symmetric $\EE$ collisions provided by the BEPCII storage ring~\cite{BEPCII} in the c.m.~energy range from 2.00 up to 4.95~GeV, with a peak luminosity of $1.1\times10^{33}$ cm$^{-2}$s$^{-1}$ achieved at $\sqrt{s} = 3.773$~GeV.
The cylindrical core of the BESIII detector covers 93\% of the full solid angle and consists of a helium-based multilayer drift chamber (MDC), a plastic scintillator time-of-flight system (TOF), and a CsI(Tl) electromagnetic calorimeter (EMC), which are all enclosed in a superconducting solenoidal magnet providing a 1.0~T magnetic field~\cite{BESIII_Detec_new}.
The solenoid is supported by an octagonal flux-return yoke with resistive plate counter muon identification modules interleaved with steel.
The charged particle momentum resolution at 1~GeV$/c$ is 0.5\%, and the d$E/$d$x$ resolution is 6\% for electrons from Bhabha scattering.
The EMC measures photon energies with a resolution of 2.5\% (5\%) at 1~GeV in the barrel (end cap) region.
The time resolution in the TOF barrel region is 68~ps, while that in the end cap region is 110~ps.

Simulated data samples produced with a {\sc geant4}-based~\cite{geant} Monte Carlo (MC) package, which includes the geometric description of the BESIII detector and the detector response, are used to determine detection efficiencies and to estimate backgrounds.
To estimate the detection efficiency, signal MC events are generated by \textsc{ConExc}~\cite{ConExc} using an amplitude model with all parameters fixed to the results obtained from the Feynman Diagram Calculation-Partial Wave Analysis (FDC-PWA) package~\cite{fdc}.
Inclusive hadronic MC samples are generated with a hybrid generator that integrates \textsc{ConExc}~\cite{ConExc}, \textsc{PHOKHARA}~\cite{phokhara} and \textsc{LUARLW}~\cite{luarlw}.

\section{EVENT SELECTION}

Candidate $e^{+}e^{-}\to p\bar{p}\pi^{0}$ events are reconstructed using the final state $p\bar{p}\gamma\gamma$.
Charged tracks detected in the MDC are required to be within a polar angle~($\theta$) range of $\left|\cos\theta\right|<0.93$, where $\theta$ is defined with respect to the $z$ axis, which is the symmetry axis of the MDC.
The distance of closest approach to the interaction point~(IP) must be less than 10\,cm along the $z$ axis (denoted as $|V_{z}|$) and less than 1\,cm in the transverse plane (denoted as $|V_{xy}|$).
For each signal candidate, one or two charged tracks are required, in the case of partial or full reconstruction, respectively, as explained in the next subsections.
Particle identification~(PID) for charged tracks combines measurements of the energy deposited in the MDC~(d$E$/d$x$) and the flight time in the TOF to calculate the probabilities for the pion, kaon, and proton hypotheses.
In particular, for partial reconstruction the probability for the electron hypothesis is also considered.
The particle type with the highest probability is assigned to the charged track.
Charged tracks are constrained to originate from a common vertex.

Photon candidates are identified using showers in the EMC.
The deposited energy of each shower must be more than 25~MeV in the barrel region~($\left|\cos\theta\right|<0.80$) and more than 50~MeV in the end cap region~($0.86<\left|\cos\theta\right|<0.92$).
To suppress electronic noise and showers unrelated to the event, the difference between the EMC time and the event start time is required to be within [0, 700]\,ns.

At low center-of-mass energies near the $p\bar{p}\pi^0$ threshold, the detection efficiency decreases due to low momenta of the pions and (anti-)protons, which are less than 0.2 GeV$/c$ and 0.35 GeV$/c$ at $\sqrt{s}=2.1250$~GeV, respectively.
At higher $\sqrt{s}$, from 2.3094 to 3.0800~GeV, the increased momenta of the pions and (anti-)protons improve the detection efficiency.
Therefore, we use a partial reconstruction technique at $\sqrt{s}$ from 2.1000 to 2.2324~GeV, and employ full reconstruction from 2.3094 to 3.0800~GeV.

\subsection{Partial Reconstruction \label{partialrec}}

Partial reconstruction of signal candidates is categorized into two modes.  
In Mode~I, events with a missing $\pi^0$ are reconstructed; while in Mode~II, events are missing a proton~(${p}_{\text{miss}}$). 
In Mode~II, the $\pi^0$ is reconstructed by two photons.
Both modes are used to increase the efficiency, and they are mutually exclusive with no overlap.

In Mode~I, one proton, one anti-proton, and no photons are required.
In order to suppress cosmic ray and beam backgrounds, the angle between the two charged tracks~(${\theta}_{p\bar{p}}$) is required to be less than 173$^{\circ}$.
A requirement on the recoil polar angle of the $p\bar{p}$ system~($\cos{\theta}_{p\bar{p}}^{\text{recoil}}$), $\left|\cos{\theta}_{p\bar{p}}^{\text{recoil}}\right|\leq0.96$, is used to suppress the dominant background process, $e^{+}e^{-}\to p\bar{p}$ produced along with initial state radiation~(ISR).
The signal yield is obtained by performing an unbinned maximum-likelihood fit to the recoil mass spectrum of the $p\bar{p}$ pair~($ M_{p\bar{p}}^{\text{recoil}}$).
The signal is modeled using the MC-simulated shape convolved with a Gaussian function to account for potential differences in calibration and resolution between data and MC simulation.
The background is modeled using a Landau function~\cite{landau}.
The fit result for data taken at $\sqrt{s}=2.1250$~GeV is shown in Fig.~\ref{fig:fit}.
An identical set of event selection criteria and the same fit procedure are applied for all data below $\sqrt{s}=2.2324$~GeV.  The fit results~($N_{\text{Signal}}^{\text{Mode~I}}$) are summarized in Table~\ref{tab:missSY}.

In Mode~II, at least two photons, one anti-proton, and no protons are required.
To exclude showers that originate from the anti-proton, the angle subtended by the EMC shower and the position of the closest anti-proton track at the EMC must be greater than ${30}^{\circ}$ as measured from the IP~(${\theta}_{\gamma\bar{p}}$).
In order to further suppress background, a one-constraint~(1C) kinematic fit imposing energy-momentum conservation is performed under the hypothesis of ${p}_{\text{miss}}\bar{p}\gamma\gamma$, treating the proton as a missing particle.
The combination with the minimum ${\chi}^{2}_{\text{1C}}$ is retained, where ${\chi}^{2}_{\text{1C}}$ is the chi-square value of the 1C kinematic fit. The requirement on ${\chi}^{2}_{\text{1C}}$ has been optimized, and a cut of ${\chi}^{2}_{\text{1C}} \leq 60$ has been imposed.
The number of signal events is extracted by a simultaneous fit to the reconstructed $\pi^0$ mass spectrum, $M(\gamma\gamma)$, at each energy point below $\sqrt{s}=2.2324$~GeV. 
An example is shown in Fig.~\ref{fig:fit} for data at $\sqrt{s}=2.1250$~GeV.
The signal is modeled using the MC-simulated shape convolved with a Gaussian function.
The parameters of the smeared Gaussian function are considered as common parameters and are shared for different data samples at $\sqrt{s}$ from 2.1000 to 2.2324~GeV.
The background is described by a linear function.
The fit results~($N_{\text{Signal}}^{\text{Mode~II}}$) are listed in Table~\ref{tab:missSY}, the last column of which lists the combined results~($N_{\text{Signal}}$).
The combined results are calculated by $N_{\text{Signal}}^{\text{Mode~I}}+N_{\text{Signal}}^{\text{Mode~II}}$, with uncertainties determined through error propagation.
\begin{table}[!htb]
  \centering
  \caption{The signal yields of Mode~I~($N_{\text{Signal}}^{\text{Mode~I}}$) and Mode~II~($N_{\text{Signal}}^{\text{Mode~II}}$) and the combined results~($N_{\text{Signal}}$) at $\sqrt{s}$ from 2.1000 to 2.2324~GeV. The uncertainties are statistical only. \label{tab:missSY}}
  \begin{ruledtabular}
  \begin{tabular}{cccc}
    $\sqrt{s}$~(GeV) & $N_{\text{Signal}}^{\text{Mode~I}}$ & $N_{\text{Signal}}^{\text{Mode~II}}$ & $N_{\text{Signal}}$ \\    \hline
    2.1000           & ~{5.5}$_{-2.9}^{+3.6}$      & ~2.4$_{-1.6}^{+2.3}$         & ~7.9$_{-3.3}^{+4.3}$  \\
    2.1250           & 189.0 $\pm$ 15.8            & 105.7 $\pm$ 11.9             & 294.7 $\pm$ 19.8     \\
    2.1500           & 18.8 $\pm$ 4.6              & ~4.4$_{-2.6}^{+3.4}$         & 23.2$_{-5.3}^{+5.7}$  \\
    2.1750           & ~~78.9 $\pm$ 10.3           & 22.0 $\pm$ 5.5               & 100.9 $\pm$ 11.7      \\
    2.2200           & 163.5 $\pm$ 13.7            & 35.3 $\pm$ 6.9               & 198.8 $\pm$ 15.3     \\
    2.2324           & 175.2 $\pm$ 14.3            & 27.3 $\pm$ 6.1               & 202.5 $\pm$ 15.5     \\
  \end{tabular}
\end{ruledtabular}
\end{table}

\subsection{Full Reconstruction \label{fullrec}}

At c.m. energies above 2.3094~GeV, full reconstruction is used, where each signal candidate is reconstructed by two charged tracks and a minimum of two photons.
One proton and one anti-proton with zero net charge are required.
The angle subtended by the EMC shower and the position of the closest charged proton track at the EMC must be greater than ${10}^{\circ}$ as measured from the IP.  The corresponding requirement for the anti-proton differs from the one adopted for Mode~II and has been optimized at ${20}^{\circ}$ instead of ${30}^{\circ}$.
A four-constraint (4C) kinematic fit, which ensures energy-momentum conservation for a $p\bar{p}\gamma\gamma$ final state, is applied to the accepted candidates.
In events containing more than two photon candidates, the combination with the minimum chi-square of the 4C kinematic fit~($\chi^2_{\text{4C}}$) is selected.
Candidate events with ${\chi}^2_{\rm 4C}\leq100$ are kept for further analysis.

\begin{figure*}[!htb]
	\mbox
	{
		\begin{overpic}[width=2.1in]{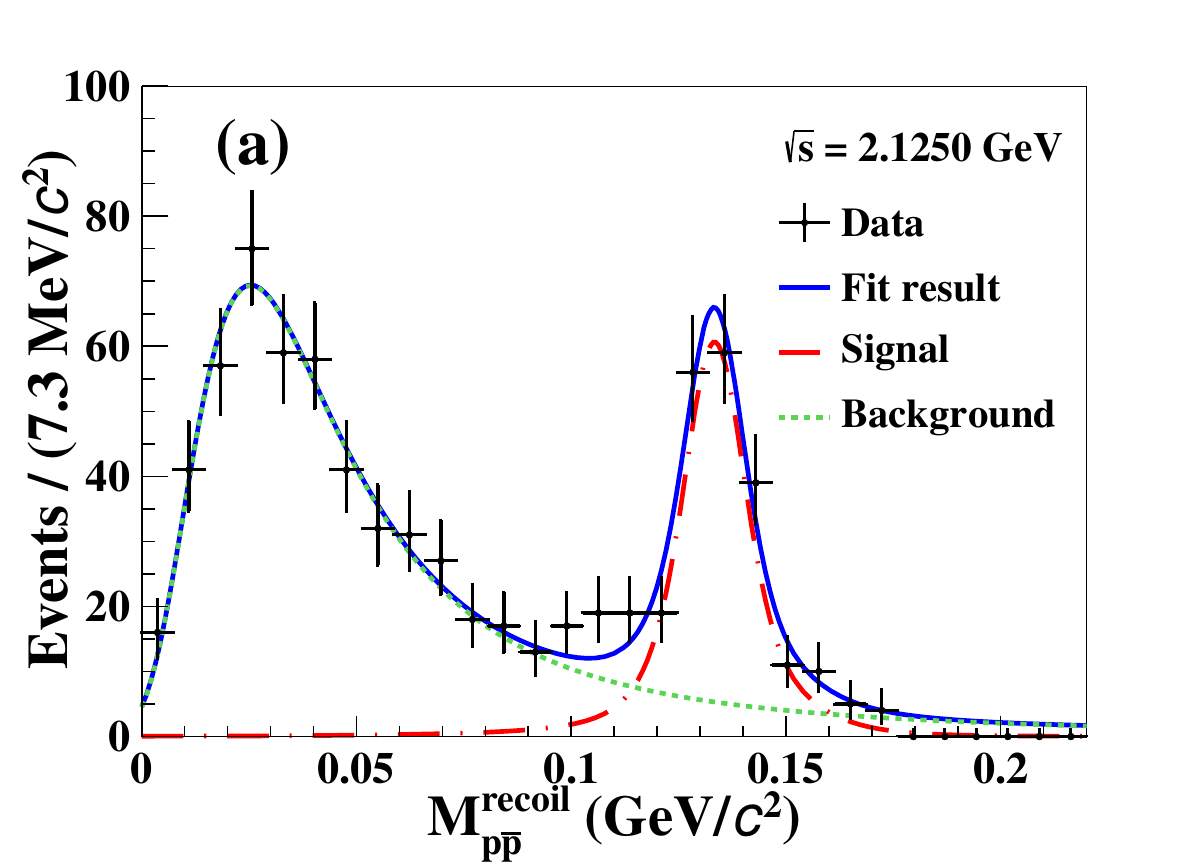}
		\end{overpic}
	}
	{
		\begin{overpic}[width=2.1in]{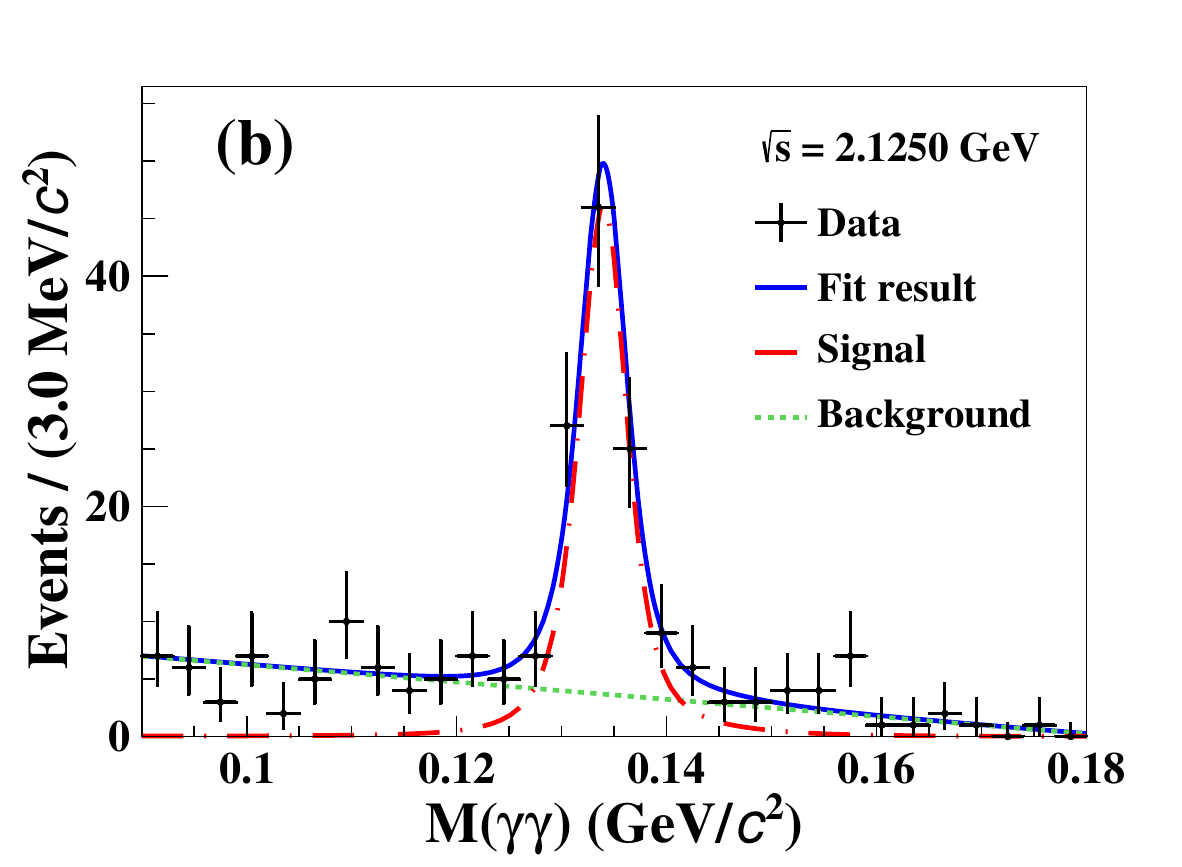}
		\end{overpic}
	}
	{
		\begin{overpic}[width=2.1in]{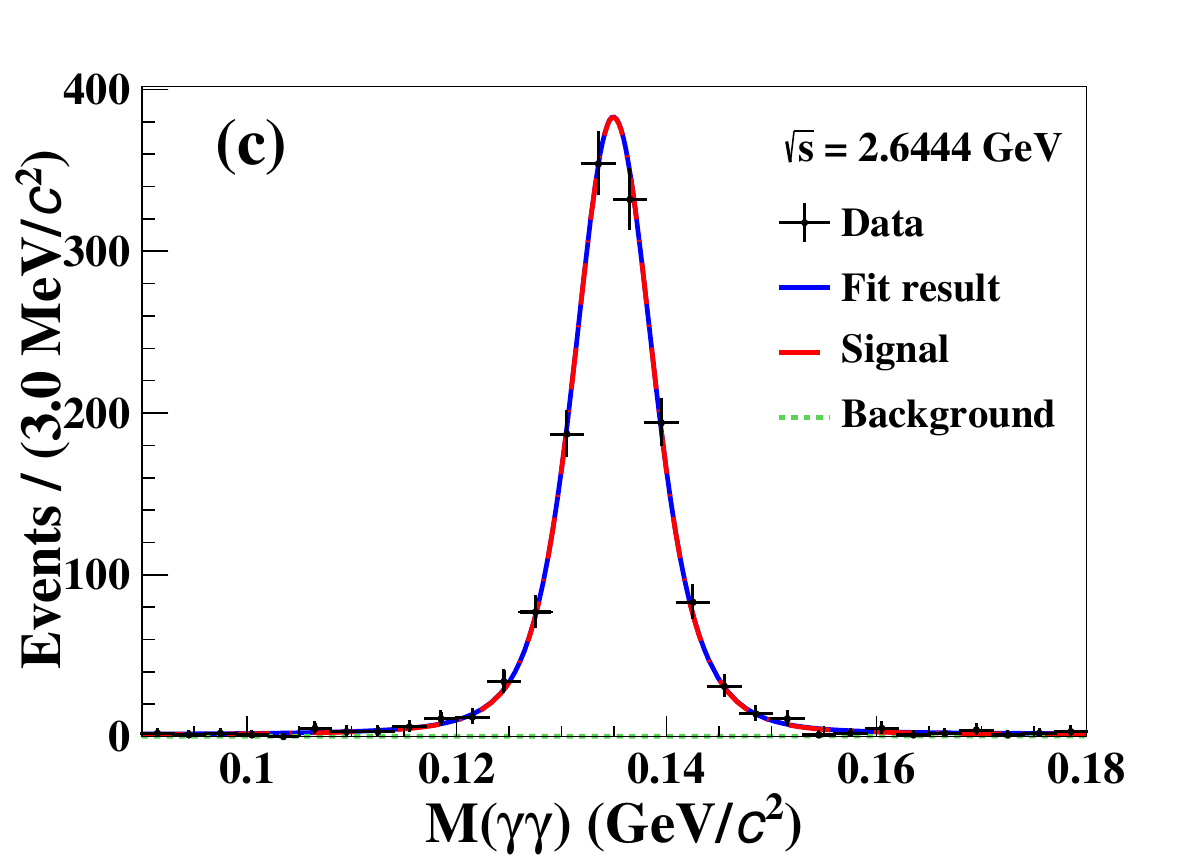}
		\end{overpic}
	}
	\caption{Fits to the distributions of (a)~$M_{p\bar{p}}^{\text{recoil}}$ in Mode~I and (b)~$M(\gamma\gamma)$ in Mode~II at $\sqrt{s}=2.1250$~GeV, as well as (c)~$M(\gamma\gamma)$ at $\sqrt{s}=2.6444$~GeV for full reconstruction. The black dots with error bars are data and the blue solid lines are the fit results. The red dash-dotted lines are the fitted signal shapes, while the green dashed lines are the fitted background shapes. \label{fig:fit}}
\end{figure*}

After all the above selection criteria are applied, the signal yield for the $e^{+}e^{-}\to p\bar{p}{\pi}^{0}$ process is obtained by fitting the $M(\gamma\gamma)$ spectrum.
The signal shape is derived from the signal MC sample and is convolved with a Gaussian function, where the Gaussian parameters are float at different energy points from 2.3094 to 3.0800~GeV.
A linear function is used to describe the background.
Figure~\ref{fig:fit} illustrates the fit result at $\sqrt{s}=2.6444$~GeV as an example and fit results for all energy points are summarized in Table~\ref{tab:crosec}.
\begin{table*}[!htb]
  \centering
  \caption{The measured Born cross sections for $\EE\ar p\bar{p}{\pi}^{0}$, where the first uncertainties are statistical and the second ones are systematic. At $\sqrt{s}$ from 2.1000 to 2.2324~GeV, partial reconstruction is used. At $\sqrt{s}$ from 2.3094 to 3.0800~GeV, full reconstruction is used and the Born cross sections have been corrected by $\mathcal{B}({\pi}^{0}\to \gamma\gamma)$. \label{tab:crosec}}
  \begin{ruledtabular}
    \begin{tabular}{c l c c c c r}
      $\sqrt{s}$~(GeV) & ~~$\mathcal{L}$~($\rm{{pb}^{-1}}$) & $N_{\text{Signal}}$ & $\epsilon$~(\%) & $(1+\delta)$ & $\frac{1}{{|1-\Pi|}^{2}}$ & $\sigma^{\rm B}$~(pb)\quad\quad\quad \\
      \hline
      2.1000  & $12.2  \pm  0.1 $ & $~~~{7.9}_{-3.3}^{+4.3} $& 8.7  & 0.794 & 1.039 & $\quad~{9.0}_{-3.8}^{+4.9}~~~~\pm 1.2 $  \\
      2.1250  & $~108  \pm  1   $ & $~~294.7  \pm  19.8$     & 17.6 & 0.804 & 1.039 & $~~18.6  \pm ~1.2~  \pm 0.8$  \\
      2.1500  & $2.84  \pm  0.02$ & $~~{23.2}_{-5.3}^{+5.7}$ & 25.3 & 0.816 & 1.040 & $~~~{38.1}_{-8.7}^{+9.4}~~~~\pm 2.0 $ \\
      2.1750  & $10.6  \pm  0.1 $ & $~~100.9  \pm  11.7$     & 31.8 & 0.828 & 1.040 & $~~34.8  \pm ~4.0~  \pm 2.1$  \\
      2.2000  & $13.7  \pm  0.1 $ & $~~198.8  \pm  15.3$     & 32.6 & 0.845 & 1.040 & $~~50.7  \pm ~3.9~  \pm 2.5$  \\
      2.2324  & $11.9  \pm  0.1 $ & $~~202.5  \pm  15.5$     & 42.5 & 0.867 & 1.041 & $~~44.4  \pm ~3.4~  \pm 2.3$  \\
			\hline
      2.3094  & $21.1  \pm  0.1 $ & $~~300.3  \pm  17.9$     & 23.5 & 0.876 & 1.041 & $~~67.2  \pm ~4.0~  \pm 2.5$  \\
      2.3864  & $22.5  \pm  0.2 $ & $~~568.3  \pm  25.2$     & 26.6 & 0.866 & 1.041 & $106.5   \pm ~4.7~  \pm 3.9$ \\
      2.3960  & $66.9  \pm  0.5 $ & $1824.7   \pm  44.0$     & 26.4 & 0.866 & 1.041 & $115.9   \pm ~2.8~  \pm 4.2$ \\
      2.5000  & $1.10  \pm  0.01$ & $~~34.7   \pm  6.0 $     & 29.8 & 0.899 & 1.041 & $114.5   \pm 19.8   \pm 8.7$ \\
      2.6444  & $33.7  \pm  0.2 $ & $1376.8   \pm  38.6$     & 31.5 & 0.958 & 1.039 & $131.9   \pm ~3.7~  \pm 4.1$ \\
      2.6464  & $34.0  \pm  0.3 $ & $1386.9   \pm  38.0$     & 31.7 & 0.958 & 1.039 & $130.8   \pm ~3.6~  \pm 4.3$ \\
      2.7000  & $1.04  \pm  0.01$ & $~~39.0   \pm  6.2 $     & 31.2 & 0.981 & 1.039 & $119.2   \pm 19.0   \pm 3.6$ \\
      2.8000  & $1.01  \pm  0.01$ & $~~28.0   \pm  5.3 $     & 33.8 & 1.022 & 1.037 & $~~78.3  \pm 14.8   \pm 3.3$ \\
      2.9000  & $~105  \pm  1   $ & $2576.2   \pm  52.5$     & 33.6 & 1.063 & 1.033 & $~~67.4  \pm ~1.4~  \pm 2.0$  \\
      2.9500  & $15.9  \pm  0.1 $ & $~~318.0  \pm  17.8$     & 33.4 & 1.084 & 1.029 & $~~54.3  \pm ~3.0~  \pm 1.8$  \\
      2.9810  & $16.1  \pm  0.1 $ & $~~330.0  \pm  18.2$     & 33.7 & 1.096 & 1.025 & $~~54.8  \pm ~3.0~  \pm 1.6$  \\
      3.0000  & $15.9  \pm  0.1 $ & $~~290.3  \pm  17.6$     & 33.7 & 1.103 & 1.021 & $~~48.7  \pm ~3.0~  \pm 1.6$  \\
      3.0200  & $17.3  \pm  0.1 $ & $~~321.0  \pm  17.9$     & 34.2 & 1.115 & 1.014 & $~~48.6  \pm ~2.7~  \pm 1.4$  \\
      3.0800  & $~126  \pm  1   $ & $1832.4   \pm  43.7$     & 31.4 & 1.178 & 0.915 & $~~43.4  \pm ~1.0~  \pm 1.3$  \\
    \end{tabular}
  \end{ruledtabular}
\end{table*}

\section{BORN CROSS SECTIONS OF $e^{+}e^{-}\to p\bar{p}\pi^{0}$}
The Born cross section~(${\sigma}^{\text{B}}$) for $e^{+}e^{-}\to p\bar{p}{\pi}^{0}$ at each energy point is determined by
\begin{equation} \label{crosec}
  {\sigma}^{\text{B}}(\sqrt{s}) = \frac{N_{\text{Signal}}}{\mathcal{L}\cdot\epsilon\cdot\mathcal{B}({\pi}^{0}\to \gamma\gamma)\cdot(1+\delta)\cdot\frac{1}{{|1-\Pi|}^{2}}},
\end{equation}
where $N_{\text{Signal}}$ is the number of signal events, $\mathcal{L}$ is the integrated luminosity of data, $(1+\delta)$ is the ISR correction factor and $\frac{1}{{|1-\Pi|}^{2}}$ is the vacuum polarization~(VP) factor.
For the full reconstruction, the $\pi^0$ only decays to $\gamma\gamma$ in the signal MC simulation, so the Born cross sections are corrected by the branching fraction of ${\pi}^{0}\to \gamma\gamma$~($\mathcal{B}({\pi}^{0}\to \gamma\gamma)$) given by the Particle Data Group (PDG)~\cite{pdg}.
In both Mode~I and Mode~II, $\mathcal{B}({\pi}^{0}\to \gamma\gamma)$ has been considered in the MC simulation.
Additionally, $\epsilon$ represents the detection efficiency obtained by analyzing the signal MC events.

\begin{figure*}
  \mbox
  {
    \begin{overpic}[width=2.1in]{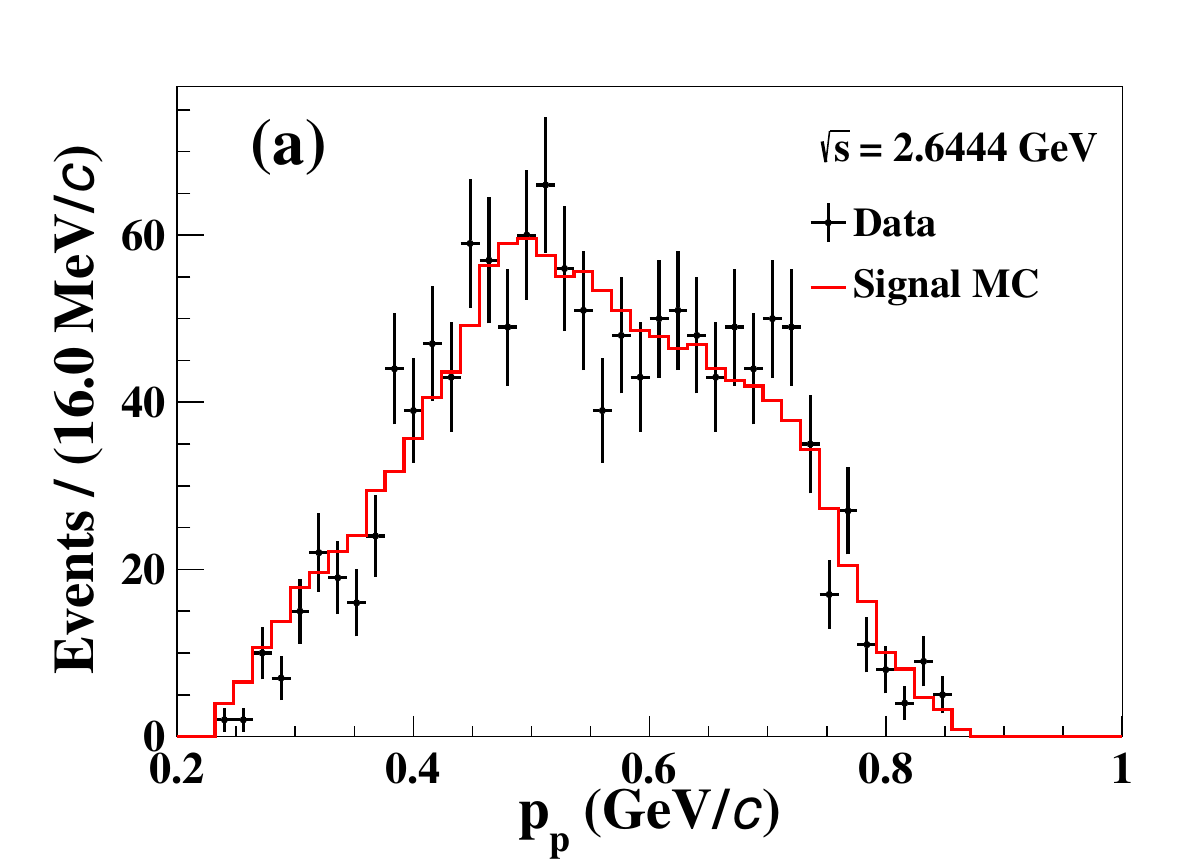}
    \end{overpic}
  }
  {
    \begin{overpic}[width=2.1in]{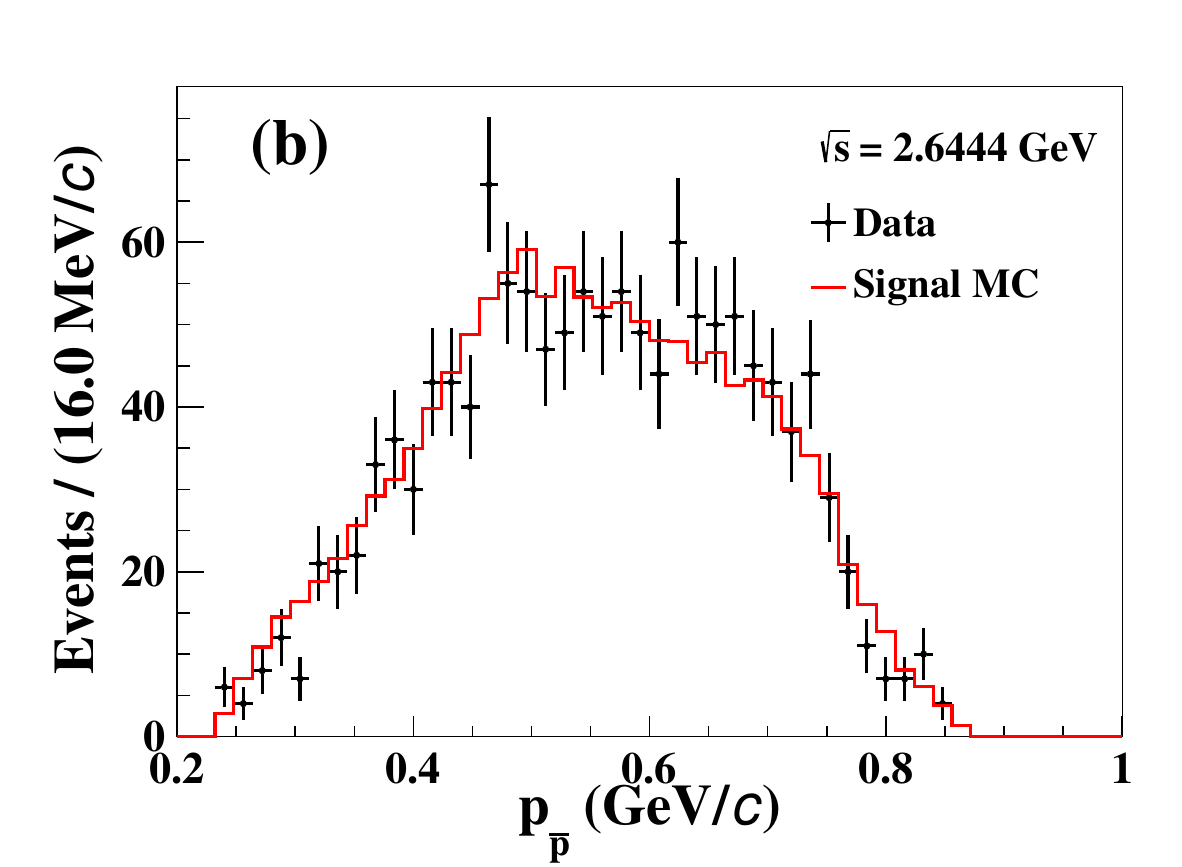}
    \end{overpic}
  }
  {
    \begin{overpic}[width=2.1in]{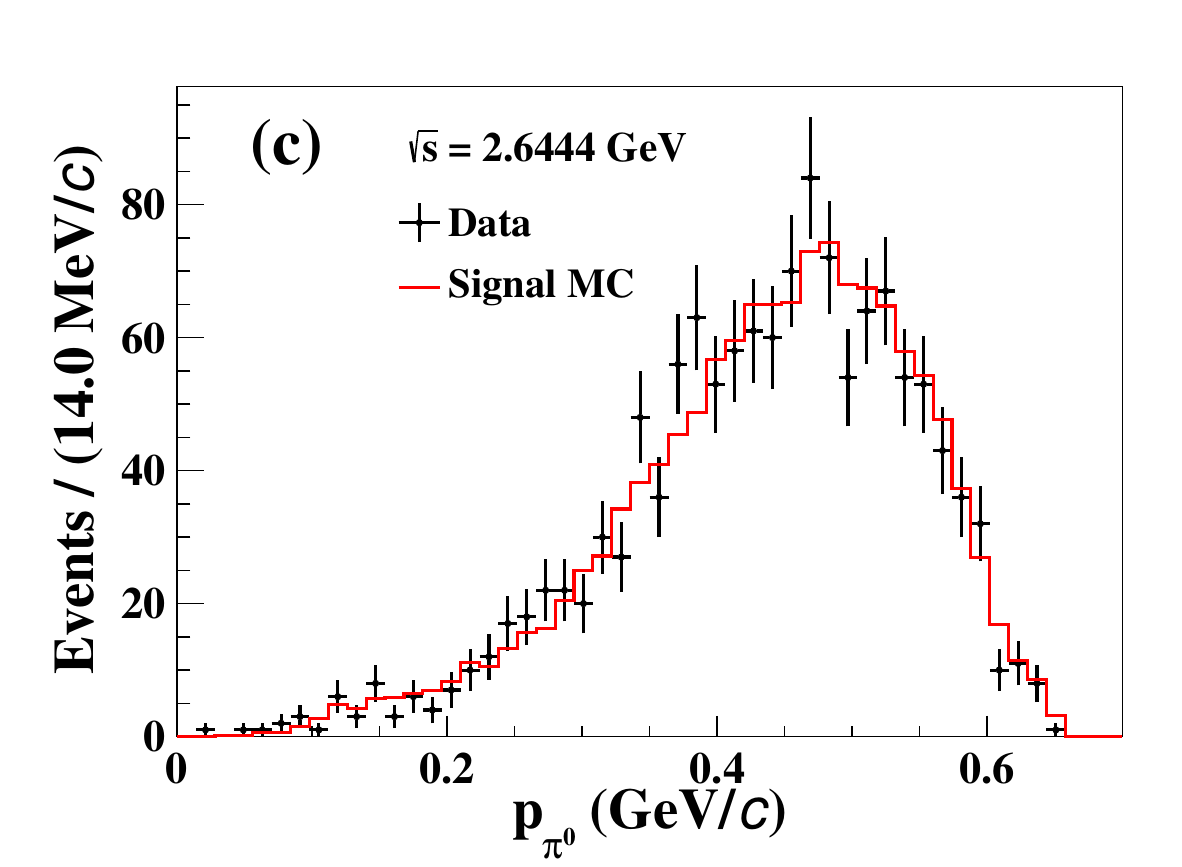}
    \end{overpic}
  }
  {
    \begin{overpic}[width=2.1in]{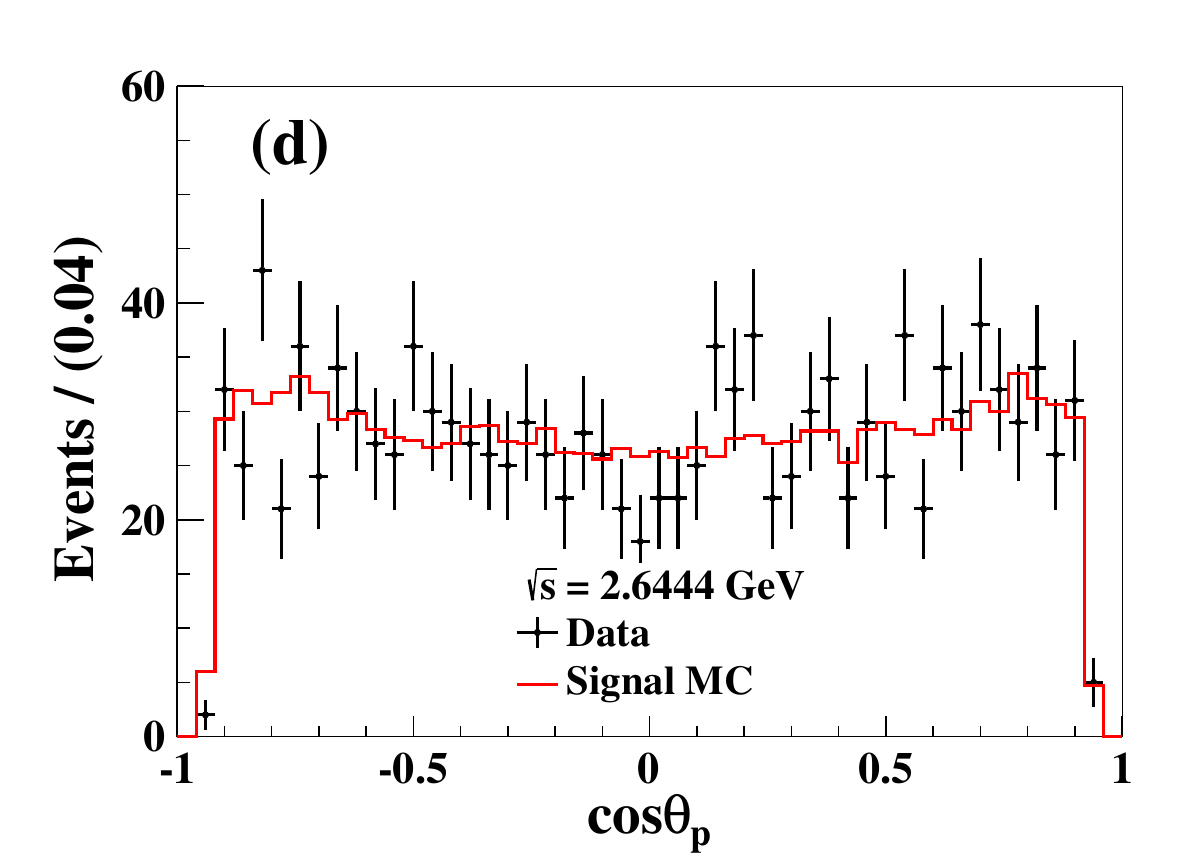}
    \end{overpic}
  }
  {
    \begin{overpic}[width=2.1in]{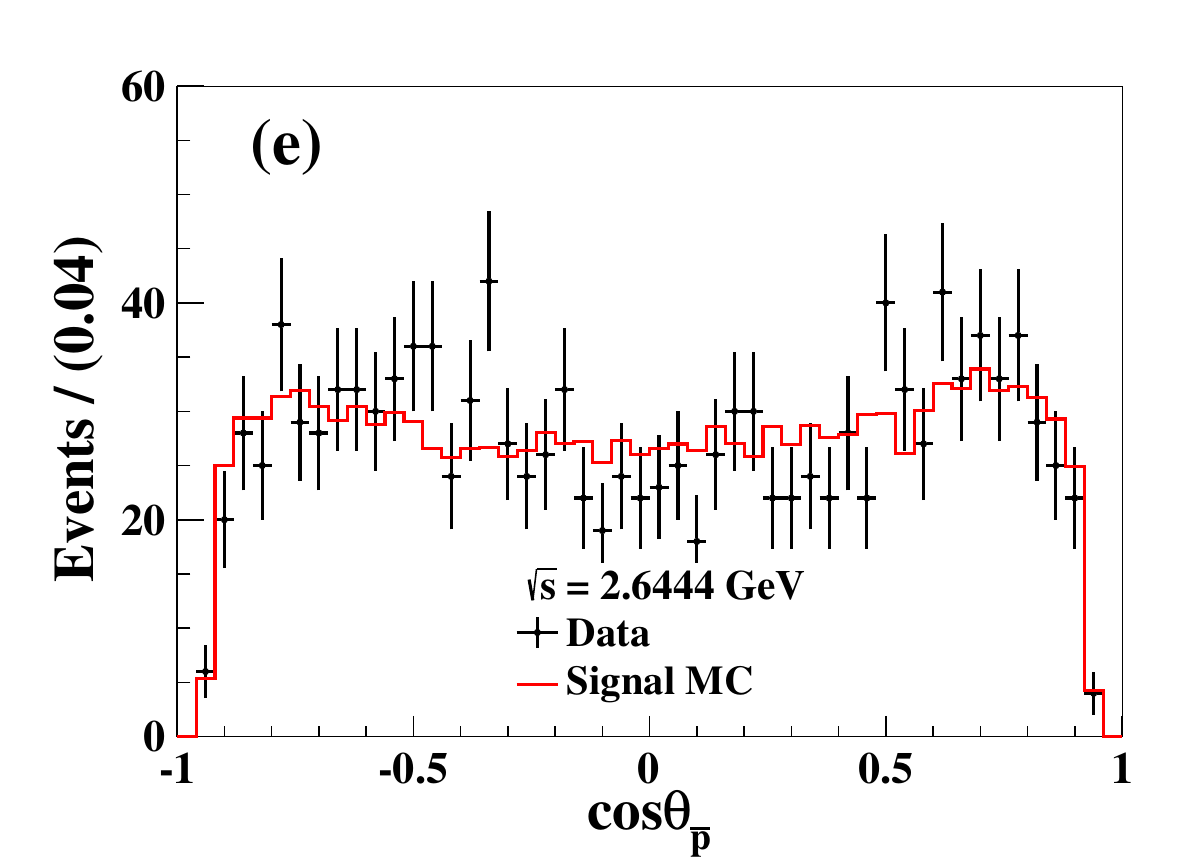}
    \end{overpic}
  }
  {
    \begin{overpic}[width=2.1in]{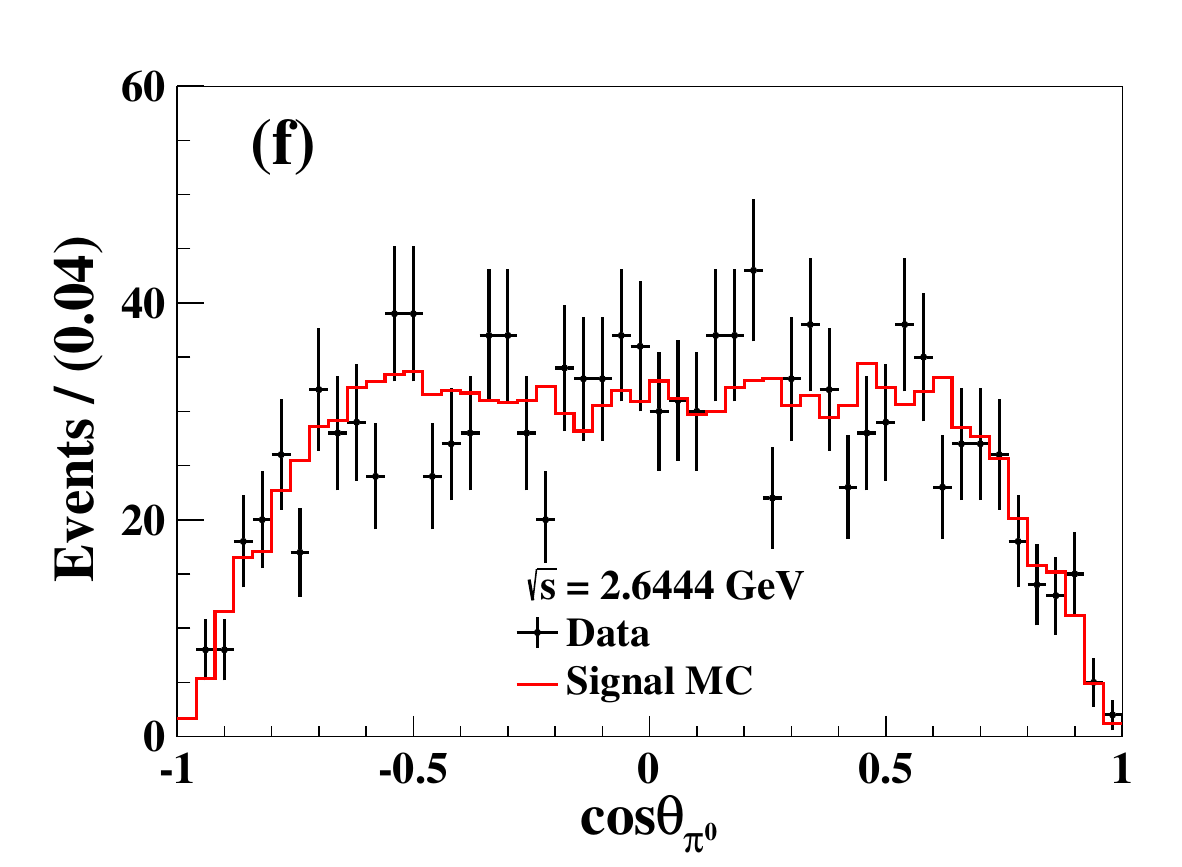}
    \end{overpic}
  }
  \caption{The momentum distributions of (a)~$p$, (b)~$\bar{p}$ and (c)~$\pi^{0}$ and the angular distributions of (d)~$p$, (e)~$\bar{p}$ and (f)~$\pi^{0}$ at $\sqrt{s}=2.6444$~GeV. The black dots with error bars are data and the red histograms are signal MC sample. \label{fig:pwaresult}}
\end{figure*}

The signal MC events are generated using an amplitude model.
The amplitudes of the quasi-two-body decays in the sequential decays are constructed using covariant tensor amplitudes~\cite{fdc}.
At $\sqrt{s}=2.6444$~GeV the process $e^{+}e^{-}\to p\bar{p}{\pi}^{0}$ is found to be well described by the following four subprocesses: $e^{+}e^{-}\to\rho(1900){\pi}^{0}$, $e^{+}e^{-}\to N(1520)p$, $e^{+}e^{-}\to N(1650)p$ and $e^{+}e^{-}\to\Delta(1232)p$, where the charge conjugation processes are also considered.
The intermediate states are parameterized by a relativistic Breit-Wigner formula, with mass and width fixed to the values from the PDG~\cite{pdg}.
The complex coupling constants of the amplitudes are determined by an unbinned maximum-likelihood fit.
The selection of components is evaluated at the energy points with the relatively high statistics, {\it i.e.}, at 2.1250, 2.3960, 2.6444 combined together with 2.6464, 2.9000 and 3.0800~GeV.
These solutions are set as reference for other data sets at nearby energy points, and the data sets at 2.1000, 2.1500, and 2.1750~GeV use the fit results obtained at 2.1250 GeV due to their limited statistics.
The solution used at each energy point is summarized in Table~\ref{tab:pwamodel}.
\begin{table*}
  \centering
  \caption{The selected components at each energy point except $\sqrt{s}=$ 2.1000, 2.1500 and 2.1750~GeV. These three energy points use the fit results obtained by $\sqrt{s}=2.1250$~GeV.  \label{tab:pwamodel}}
  \resizebox{\linewidth}{!}{
  \begin{tabular}{ll}
    \hline\hline
    ~~~~~~~~~~$\sqrt{s}$~(GeV)~~~~~~~~~~&~~~~~~~~~~Components selected~~~~~~~~\\
    \hline
    ~~~~~~~~~~2.1250,~~~2.2000,~~~2.2324~~~~~~~~~~  &  ~~~~~~~~~~$N(940)$,~~~$\rho(1900)$~~~~~~~~~~ \\
    ~~~~~~~~~~2.3094,~~~2.5000~~~~~~~~~~          &  ~~~~~~~~~~$N(940)$,~~~$\Delta(1232)$,~~~$\rho(2000)$~~~~~~~~~~ \\
    ~~~~~~~~~~2.3864~~~~~~~~~~                  &  ~~~~~~~~~~$\Delta(1232)$,~~~$\rho(2000)$~~~~~~~~~~ \\
    ~~~~~~~~~~2.3960~~~~~~~~~~                  &  ~~~~~~~~~~$N(940)$,~~~$\Delta(1232)$,~~~$\rho(2000)$,~~~$\rho(2225)$~~~~~~~~~~ \\
    ~~~~~~~~~~2.6444,~~~2.6464,~~~2.7000~~~~~~~~~~  &  ~~~~~~~~~~$\Delta(1232)$,~~~$N(1520)$,~~~$N(1650)$,~~~$\rho(1900)$~~~~~~~~~~ \\
    ~~~~~~~~~~2.8000~~~~~~~~~~                  &  ~~~~~~~~~~$N(940)$,~~~$\Delta(1232)$,~~~$\rho(2225)$~~~~~~~~~~ \\
    ~~~~~~~~~~2.9000,~~~2.9500,~~~2.9810~~~~~~~~~~  &  ~~~~~~~~~~$\Delta(1232)$,~~~$N(1440)$,~~~$N(1520)$,~~~$N(1710)$,~~~$\rho(2000)$,~~~$\rho(2225)$~~~~~~~~~~ \\
    ~~~~~~~~~~3.0000,~~~3.0200,~~~3.0800~~~~~~~~~~  &  ~~~~~~~~~~$\Delta(1232)$,~~~$\Delta(1600)$,~~~$N(1520)$,~~~$N(1710)$,~~~$N(1895)$~~~~~~~~~~ \\
    \hline\hline
  \end{tabular}
}
\end{table*}
At all c.m. energies, the weighted signal MC simulation is in reasonable agreement with data, which can be seen at $\sqrt{s}=2.6444$~GeV in Fig.~\ref{fig:pwaresult} as an example.

\begin{figure}[!htb]
  \includegraphics[width=3in]{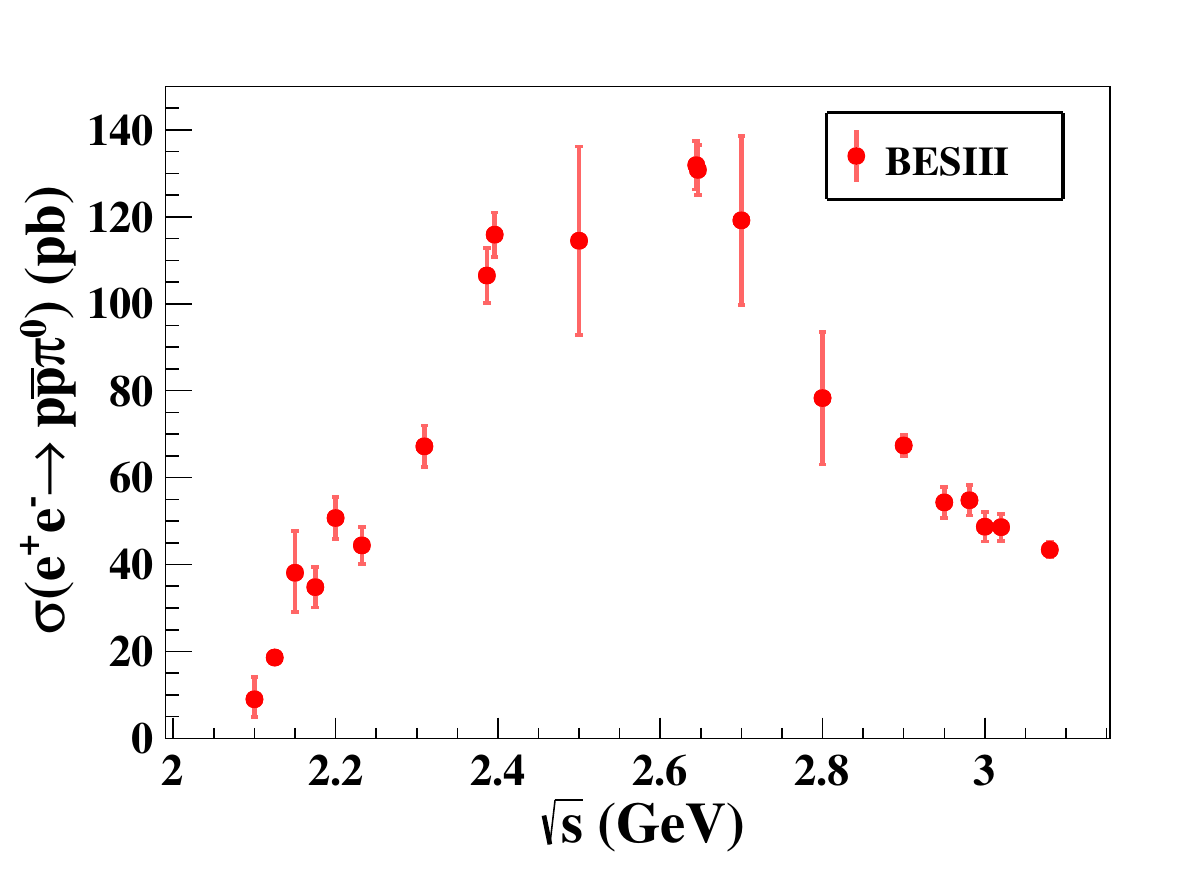}  
  \caption{The measured Born cross sections for the process $\EE\ar p\bar{p}{\pi}^{0}$.
  \label{fig:crossec}}
\end{figure}

The ISR correction factor and the detection efficiency are estimated based on signal MC samples, and then weighted by an iterative MC weighting method~\cite{ISRiteration}, with the iterative procedure repeated until the measured Born cross sections converge.
The corresponding Born cross sections are summarized in Table~\ref{tab:crosec} and shown in Fig.~\ref{fig:crossec}.

\section{SYSTEMATIC UNCERTAINTY}
Several sources of systematic uncertainties are considered in the determination of the Born cross sections.
The uncertainty of the luminosity measurement is less than 0.9\%~\cite{2125lum}.
The uncertainties associated with the tracking and PID efficiency for (anti-)protons are studied by a control sample of $\jpsi\ar p\bar{p}{\pi}^{+}{\pi}^{-}$ and are obtained by averaging the uncertainties weighted by signal events in each transverse momentum bin.
The uncertainty due to the photon reconstruction is 1.0\%, leading to a total uncertainty of 2.0\% for the two photons for Mode~II~\cite{photonsys}.
Both uncertainties due to the ${\theta}_{\gamma\bar{p}}\geq 30^{\circ}$ and ${\theta}_{\gamma\bar{p}}\geq 20^{\circ}$ requirements are studied by a control sample of $\jpsi\ar p\bar{p}{\pi}^{0}$, which are set to be 0.8\% and 0.3\% per photon for Mode~II and full reconstruction, respectively.
The uncertainty of the branching fraction of $\pi^0\to\gamma\gamma$ quoted from the PDG~\cite{pdg} is 0.03\% for full reconstruction, which is neglected.

In Mode~I, the uncertainties due to both ${\theta}_{p\bar{p}}\le 173^{\circ}$ and $|\cos\theta_{{\pi}^{0}}|\le 0.96$ requirements are assigned as 0.3\% and 0.4\% based on a study of the $J/\psi\to p\bar{p}{\pi}^{0}$ control sample.
In Mode~II, the uncertainty from the kinematic fit is also studied with the $J/\psi\to p\bar{p}{\pi}^{0}$ control sample. 
The difference in selection efficiency before and after the kinematic fit for data and MC simulation, 0.1\%, is taken as the systematic uncertainty.
In full reconstruction, the uncertainty due to the $\pi^0$ reconstruction, including the photon selection efficiency, is estimated to be 2.5\%, based on an analysis of the $\jpsi\ar p\bar{p}{\pi}^{0}$ control sample.

The uncertainty from the ISR and VP correction factors is assigned to be 0.5\% according to the accuracy of the radiation function~\cite{radiationfunc}.
The additional contribution from the cross section line shape is evaluated by varying the parameters of the fitting model.
All parameters are randomly varied within their uncertainties and the resulting parametrization of the line shape is used to recalculate $(1+\delta)$, $\epsilon$ and the corresponding cross sections.
This procedure is repeated 1000 times and the standard deviation of the resulting cross sections is taken as the systematic uncertainty.

The uncertainties related to the fit procedure are estimated by changing the fit range, replacing the signal shape with a Breit-Wigner convolved with a Gaussian function, and replacing the background shape with a second-order polynomial function except in Mode~I whose background shape is changed to the MC-simulated shape from the exclusive process $e^{+}e^{-}\to p\bar{p}$.
This MC sample of $e^{+}e^{-}\to p\bar{p}$ is generated by \textsc{PHOKHARA}~\cite{phokhara} with both ISR and final state radiation enabled.
The uncertainties due to the background shape for full reconstruction are neglected.
The uncertainty arising from the signal MC model is evaluated using an alternative amplitude model, while the quoted mass and width are smeared with their associated uncertainties.
The relative difference in the detection efficiency is taken as the uncertainty.

At $\sqrt{s}$ from 2.1000 and 2.2324~GeV two partial reconstruction methods~(Mode~I and Mode~II) are applied.
The uncertainties of these two methods are weighted by detection efficiencies and averaged. The weighted average formula is
\begin{equation} \label{ave_sys}
  \sigma_{\rm tot}^{2}=\sum_{i=1}^{2}\omega_{i}^{2}\sigma_{i}^{2}+\sum_{i,j=1;i\neq j}^{2}\rho_{ij}\omega_{i}\omega_{j}\sigma_{i}\sigma_{j},
\end{equation}
with
\begin{equation} \label{weight}
  \omega_{i}=\frac{\varepsilon_{i}}{\sum_{i=1}^{2}\varepsilon_{i}},
\end{equation}
where $\omega_{i}$, $\sigma_{i}$ and $\varepsilon_{i}$ with $i = 1, 2$ are the weight, systematic uncertainty and efficiency for reconstruction method $i$, respectively, and $\rho_{ij}$ is the correlation parameter for the two different methods $i$ and $j$.
For the uncorrelated systematic uncertainties, like the uncertainties from the photon reconstruction efficiency, the requirements of ${\theta}_{\gamma\bar{p}}\geq 30^{\circ}$, ${\theta}_{p\bar{p}}\le 173^{\circ}$ and $|\cos\theta_{{\pi}^{0}}|\le 0.96$, the kinematic fit and fit procedure, the $\rho_{ij}$ values are set to be zero.
For other correlated systematic uncertainties, like the uncertainties related to luminosity, tracking, PID, ISR and VP correction factors and the MC model, the $\rho_{ij}$ are set to be one.
After weighting, the systematic uncertainty of the kinematic fit is less than 0.1\% and is neglected.

The combined results from different partial reconstruction methods are summarized in Table~\ref{tab:sys_low}, while the systematic uncertainties related to full reconstruction are listed in Table~\ref{tab:sys_high}.
The total systematic uncertainty is summed in quadrature.

\begin{table*}[!htb]
  \centering
  \caption{Relative systematic uncertainties~(in \%) on the cross section measurement for each c.m. energy from 2.1000 to 2.2324~GeV. They are denoted as luminosity~($\mathcal{L}$), Tracking~(Track), PID, photon reconstruction~(Photon), requirements of ${\theta}_{\gamma\bar{p}}\geq30^{\circ}$~($\theta_{\gamma\bar{p}}$1), ${\theta}_{p\bar{p}}\le 173^{\circ}$, and $\left|\cos\theta_{{\pi}^{0}}\right|\le 0.96$, ISR correction, signal shape~(Sig), background shape~(Bkg), fit range~(Fit) and signal MC model~(MC). The last column is the total systematic uncertainty. \label{tab:sys_low}}
  \begin{ruledtabular}
    \begin{tabular}{ccccccccccccccc}
      $\sqrt{s}$~(GeV) & $\mathcal{L}$ & Track & PID & Photon & $\theta_{\gamma\bar{p}}$1 & $\theta_{p\bar{p}}$ & $\cos{\theta}_{{\pi}^{0}}$ & ISR correction & Sig & Bkg & Fit & MC & Total \\
      \hline
      2.1000  & 0.7 & 3.3 & 0.8 & 0.6 & 0.4 & 0.3 & 0.2 & 0.6 & 4.2 & 12.0 & 3.0  & 0.6 & 13.6 \\
      2.1250  & 0.8 & 3.4 & 0.8 & 0.4 & 0.3 & 0.3 & 0.3 & 0.6 & 1.3 & 1.2  & 0.4  & 1.1 & 4.2  \\
      2.1500  & 0.8 & 3.1 & 0.9 & 0.3 & 0.3 & 0.3 & 0.3 & 0.6 & 2.3 & 1.5  & 2.3  & 1.5 & 5.2  \\
      2.1750  & 0.8 & 2.9 & 0.9 & 0.3 & 0.2 & 0.3 & 0.3 & 0.6 & 4.6 & 0.3  & 1.8  & 1.6 & 6.1  \\
      2.2000  & 0.7 & 2.6 & 1.0 & 0.3 & 0.2 & 0.3 & 0.3 & 0.6 & 3.0 & 0.6  & 0.9  & 2.5 & 5.0  \\
      2.2324  & 0.7 & 2.3 & 1.0 & 0.2 & 0.2 & 0.3 & 0.3 & 0.9 & 2.9 & 1.2  & 0.4  & 2.8 & 5.1  \\
    \end{tabular}
  \end{ruledtabular}
\end{table*}
\begin{table*}[!htb]
  \centering
	\caption{Relative systematic uncertainties~(in \%) on the cross section measurement for each c.m. energy from 2.3094 to 3.0800~GeV. They are denoted as luminosity~($\mathcal{L}$), Tracking~(Track), PID, requirement of ${\theta}_{\gamma\bar{p}}\geq20^{\circ}$~($\theta_{\gamma\bar{p}}$2), $\pi^0$ reconstruction, ISR correction, signal shape~(Sig), fit range~(Fit) and signal MC model~(MC). The last column is the total systematic uncertainty. \label{tab:sys_high}}
  \begin{ruledtabular}
    \begin{tabular}{cccccccccccc}
			$\sqrt{s}$~(GeV) & $\mathcal{L}$ & Track & PID & $\theta_{\gamma\bar{p}}$2 & $\pi^0$ reconstruction & ISR correction & Sig & Fit & MC & Total \\
      \hline
      2.3094  & 0.7 & 1.9 & 1.1 & 0.6 & 2.5 & 0.5 & 0.9 & 0.3 & 0.8 & 3.7\\
      2.3864  & 0.8 & 1.9 & 1.1 & 0.6 & 2.5 & 0.5 & 1.1 & 0.4 & 0.3 & 3.7\\
      2.3960  & 0.7 & 1.8 & 1.1 & 0.6 & 2.5 & 0.5 & 0.2 & 0.3 & 1.1 & 3.6\\
      2.5000  & 0.8 & 1.6 & 0.9 & 0.6 & 2.5 & 0.5 & 6.7 & 1.2 & 0.2 & 7.6\\
      2.6444  & 0.6 & 1.3 & 0.8 & 0.6 & 2.5 & 0.5 & 0.3 & 0.1 & 0.1 & 3.1\\
      2.6464  & 0.8 & 1.3 & 0.8 & 0.6 & 2.5 & 0.5 & 1.1 & 0.1 & 0.1 & 3.3\\
      2.7000  & 0.7 & 1.1 & 0.7 & 0.6 & 2.5 & 0.5 & 0.0 & 0.6 & 0.1 & 3.1\\
      2.8000  & 0.7 & 0.8 & 0.6 & 0.6 & 2.5 & 0.5 & 0.0 & 3.1 & 0.3 & 4.2\\
      2.9000  & 0.9 & 0.8 & 0.6 & 0.6 & 2.5 & 0.5 & 0.2 & 0.2 & 0.1 & 2.9\\
      2.9500  & 0.9 & 0.8 & 0.5 & 0.6 & 2.5 & 0.5 & 1.6 & 0.2 & 0.1 & 3.3\\
      2.9810  & 0.6 & 0.8 & 0.5 & 0.6 & 2.5 & 0.5 & 0.8 & 0.4 & 0.2 & 3.0\\
      3.0000  & 0.7 & 0.7 & 0.5 & 0.6 & 2.5 & 0.5 & 1.5 & 0.4 & 0.2 & 3.2\\
      3.0200  & 0.7 & 0.7 & 0.5 & 0.6 & 2.5 & 0.5 & 0.2 & 0.3 & 0.4 & 2.9\\
      3.0800  & 0.7 & 0.8 & 0.5 & 0.6 & 2.5 & 0.5 & 0.6 & 0.2 & 0.4 & 3.0\\
    \end{tabular}
  \end{ruledtabular}
\end{table*}

\section{Summary and discussion}
This paper presents the measurement of the Born cross sections for the process $e^{+}e^{-}\to p\bar{p}\pi^{0}$ at 20 c.m. energies, corresponding to a total integrated luminosity of 636.8~pb$^{-1}$, in the energy range from 2.1000 to 3.0800 GeV.
The uncertainties at most energy points are less than 10\%. The dominant contribution to the uncertainties is statistical near the energy threshold and systematic at higher energies.
An increase around 2.4~GeV in the cross sections, which may be from the $pN(1440)$ threshold, requires further experimental studies.
No obvious threshold enhancement is observed in the cross sections.
Intermediate states are investigated utilizing the FDC-PWA package.
Due to low statistics, extracting the $N^{*}$ and ${\Delta}^{*}$ yields and measuring the cross sections of two-body processes is still challenging.
An improved PWA of the process ${e}^{+}{e}^{-}\to p\bar{p}{\pi}^{0}$ is highly desirable in order to study the intermediate resonances such as ${N}^{*}$ or ${\Delta}^{*}$ coupling to $p{\pi}^{0}$ and $\bar{p}{\pi}^{0}$ or ${1}^{-}$ vector states coupling to $p\bar{p}$.
These improved analyses could be achieved with the much larger data sets expected at the forthcoming advanced super tau-charm facilities~\cite{stcf, sctf}.

\begin{acknowledgments}
The BESIII Collaboration thanks the staff of BEPCII, the IHEP computing center and the supercomputing center of the University of Science and Technology of China~(USTC) for their strong support. 
This work is supported in part by National Key R\&D Program of China under Contracts Nos. 2020YFA0406400, 2020YFA0406300, 2023YFA1606000, 2023YFA1609400; National Natural Science Foundation of China (NSFC) under Contracts Nos. 11625523, 11635010, 11735014, 11835012, 11935015, 11935016, 11935018, 11961141012, 12005219, 12025502, 12035009, 12035013, 12061131003, 12122509, 12105276, 12175244, 12192260, 12192261, 12192262, 12192263, 12192264, 12192265, 12221005, 12225509, 12235017; the Chinese Academy of Sciences (CAS) Large-Scale Scientific Facility Program; the CAS Center for Excellence in Particle Physics (CCEPP); Joint Large-Scale Scientific Facility Funds of the NSFC and CAS under Contract Nos. U1832207, U2032111, U1732263, U1832103; CAS Key Research Program of Frontier Sciences under Contracts Nos. QYZDJ-SSW-SLH003, QYZDJ-SSW-SLH040; CAS Youth Team Program under Contract No. YSBR-101; 100 Talents Program of CAS; The Institute of Nuclear and Particle Physics (INPAC) and Shanghai Key Laboratory for Particle Physics and Cosmology; European Union's Horizon 2020 research and innovation programme under Marie Sklodowska-Curie grant agreement under Contract No. 894790; German Research Foundation DFG under Contracts Nos. 455635585, Collaborative Research Center CRC 1044, FOR5327, GRK 2149; Istituto Nazionale di Fisica Nucleare, Italy; Ministry of Development of Turkey under Contract No. DPT2006K-120470; National Research Foundation of Korea under Contract No. NRF-2022R1A2C1092335; National Science and Technology fund of Mongolia; National Science Research and Innovation Fund (NSRF) via the Program Management Unit for Human Resources \& Institutional Development, Research and Innovation of Thailand under Contract No. B16F640076; Polish National Science Centre under Contract No. 2019/35/O/ST2/02907; The Swedish Research Council; U. S. Department of Energy under Contract No. DE-FG02-05ER41374.
\end{acknowledgments}

\bibliography{main}

\end{document}